\documentclass[preprint2]{aastex61}

\usepackage{color}

\newcommand{\Rnom}{\hbox{$\mathcal{R}^{\rm N}_\odot$}}

\shorttitle{The Solar X-ray Limb}
\shortauthors{Battaglia et al.}

\begin{document}


\title{The Solar X-ray Limb }

\author{Marina Battaglia}
\affiliation{University of Applied Sciences and Arts Northwestern Switzerland, CH-5210 Windisch, Switzerland }
\author{Hugh~S. Hudson}
\affiliation{School of Physics and Astronomy, SUPA, University of Glasgow, Glasgow G12 8QQ, UK}
\affiliation{SSL, UC Berkeley, 7 Gauss Way, Berkeley, CA, USA}
\author{Gordon~J. Hurford}
\affiliation{SSL, UC Berkeley, 7 Gauss Way, Berkeley, CA, USA}
\author{S\"am Krucker}
\affiliation{University of Applied Sciences and Arts Northwestern Switzerland, CH-5210 Windisch, Switzerland }
\affiliation{SSL, UC Berkeley, 7 Gauss Way, Berkeley, CA, USA}
\author{Richard~A. Schwartz}
\affiliation{Catholic University of America at NASA Goddard Space Flight Center, Greenbelt, MD 20771, USA}

\correspondingauthor{Marina Battaglia}
\email{marina.battaglia@fhnw.ch}

\begin{abstract}
We describe a new technique to measure the height of the X-ray limb with observations from occulted X-ray flare sources as observed by the RHESSI  (the Reuven Ramaty High-Energy Spectroscopic Imager) satellite.
This method has model dependencies different from those present in traditional observations at optical wavelengths, which depend upon detailed modeling involving radiative transfer in a medium with complicated geometry and flows. It thus provides an independent and more rigorous measurement of the ``true'' solar radius, meaning that of the mass distribution.
RHESSI's measurement makes use of the flare X-ray source's spatial Fourier components (the visibilities), which are sensitive to the presence of the sharp edge at the lower boundary of the occulted source. 
We have found a suitable flare event for analysis, SOL2011-10-20T03:25 (M1.7), and report a first result from this
novel technique here. Using a 4-minute integration over the 3-25~keV photon energy range, we find $R_{\mathrm{X-ray}} = 960.11\ \pm\  0.15 \pm 0.29$ arcsec, at 1 AU, where the uncertainties include statistical uncertainties from the method and a systematic error.
The standard VAL-C model predicts a value of 959.94 arcsec, about 1$\sigma$ below our value. 
\end{abstract}



\section{Introduction}
We normally determine the outer boundary of the Sun by observations at optical wavelengths, essentially making use of the radial distance of the tangent ray for which the optical depth $\tau$ is unity at a standard wavelength, such as 5000 \AA. Such a definition depends upon details of the atmospheric structure and of the physics of the radiative transfer (absorption and scattering) within that structure.
The standard reference for the value, as cited by \citet{1973asqu.book.....A}, appears to be the fundamental observational work of \citet{1891AN....128..361A}, who made extensive visual observations with ``heliometers.'' 
These were used to measure the angular separation between the two ends of a given solar diameter.
The measurements gave the value 1919.26~$\pm$~0.10~arcsec.
A more recent standard was set by \citet{1998ApJ...500L.195B}, who used timing measurements of meridian transits in a special-purpose telescope operated between 1981 and 1987.
This yielded a value of 1919.359~$\pm$~0.018~arcsec for the observed near-equatorial diameter.
These angular measures for the location of the solar limb need an adjustment to provide estimates of the solar radius itself, expressed in SI units, because the observed limb lies several scale heights above the physical radius.
The current best value for the radius itself, incorporating many sources, gives R$_\odot$ = 695.658$\pm$140~Mm \citep{2008ApJ...675L..53H}.
This value has been rounded off and adopted within errors as the IAU nominal value \Rnom = 695.7 Mm \citep{2016AJ....152...41P} as one of a set of self-consistent values for solar and planetary properties.

What do such measurements mean physically?
From the point of view of stellar structure, we may wish to use such a measurement to characterize the spatial distribution of the solar material, and so we would like to interpret the opacity measurement as a radial distance from Sun center, for example.
For reference, the point in a semi-empirical atmospheric model such as VAL-C \citep{1981ApJS...45..635V}
corresponding to the nominal radial distance R$_\odot$ has optical depth $\tau_{5000\mathrm{\AA}} = 1$ as viewed radially from above.
This optical depth conventionally defines the location of the photosphere, or the solar radius.
According to the VAL-C model, the total mass of the atmosphere above that point is about $10^{-10}$~M$_\odot$, and so a sphere at this radial distance from Sun center (i.e., within R$_\odot$) contains most of the mass of the Sun.
The increase of optical depth for oblique rays means that the observed limb always lies above the conventional photosphere.
The difference amounts to several photospheric scale heights \citep[about 350 km; see][for a discussion of this point]{1998ApJ...500L.195B,2008ApJ...675L..53H}.
At longer and shorter wavelengths (relative to the ``opacity minimum'' at about 1.5~$\mu$m), the opacity generally increases, but its spectral behavior is greatly complicated by atomic and molecular transitions.
The radial distance of the apparent limb thus generally  increases as the opacity increases, slowly towards the infrared and more rapidly in the UV. 

Because of the presence of flows and oscillations on small scales, the surface of the Sun is not smooth.
The height of the limb thus depends on the dynamics of the upper solar atmosphere at a given wavelength, an interesting problem
normally described by numerical models and relevant to such matters as the chemical composition \citep[e.g.,][]{2009ARA&A..47..481A}.
Much literature \citep[see, e.g.,][]{2015ApJ...812...91R} has been devoted to characterizing the limb and its dynamics at all wavelengths.
In this paper we report a first attempt to make precise measurements of the X-ray limb, by a technique that is more directly related to the mass distribution, rather than to the opacity structure at longer wavelengths.
The method is based on locating the sharp lower edge of partially-occulted flare sources as observed by the Reuven Ramaty High-Energy Solar Spectroscopic Imager (RHESSI) spacecraft \citep{2002SoPh..210....3L}.
The limb thus observed is in \textit{absorption}, due to the photoeffect and Compton scattering. Photoelectric absorption, important below about 12~keV, depends upon the elemental abundances and the ionization state of the medium.
At higher energies Compton scattering becomes dominant.
In the scattering regime we have a particularly simple interpretation: the absorption ideally just depends upon the column density of electrons, free or bound, along the line of sight.
RHESSI is well-suited to observing the sharp occulted edge of the source because of its natural interpretation in terms of discrete spatial Fourier components \citep[][]{2002SoPh..210...61H}, as discussed for this application in Section~\ref{sec:theory}.
Note that \citet{1983ApJ...264..660W} used an analogous technique to study the limb at mm wavelengths.
In Section~\ref{sec:obs} we describe observations of a single partially-occulted flare, chosen as the best case after extensive searches. The calculation of the limb height and comparison with predictions from models are given in Section~\ref{sec:height}. 

\begin{figure*}[!t]
\includegraphics[width=15cm]{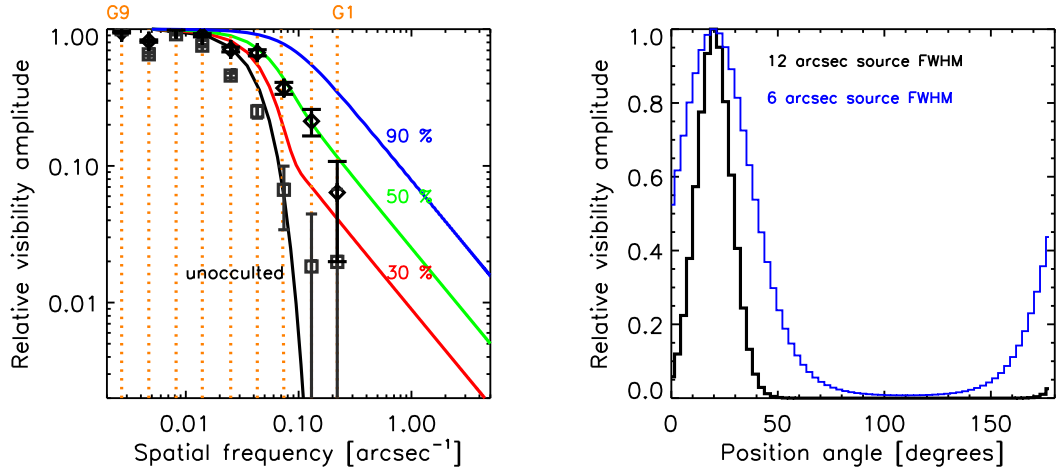}
\caption{Left: relative amplitudes of visibilities corresponding to the sharp edge of a Gaussian source of FWHM = 12 arcsec which is occulted by different fractions (given as \% in the Figure). 
The vertical dashed lines show the spatial frequencies of the RHESSI collimators. Diamonds are the relative visibility amplitudes associated with the limb signature in the event analyzed here. Rectangles give the relative amplitudes of the visibilities perpendicular to the limb.
Right: the predicted visibility amplitude vs. rotational position angle for RHESSI collimator 1 (2.26$''$ FWHM), showing a prominent peak at the angle dictated by the limb orientation in the model, and with a Gaussian width determined by the collimator resolution and source size. Note that in reality the visibility amplitude at position angles perpendicular to the source will have a finite value, unlike the idealized case, due to statistical uncertainties and source sub-structure.}
\label{fig:vis_vs_k}
\end{figure*}

\section{Visibility theory}\label{sec:theory} 
RHESSI is a collimator-based Fourier imager. The incident photon flux from a solar flare is time-modulated by two grids as the spacecraft spins around its axis. RHESSI has nine detectors and, accordingly, nine pairs of grids (G1 through G9) with different pitch-widths corresponding to angular resolutions from as small as 2.26 arcsec in the case of G1 up to 183.2 for G9 \citep{2002SoPh..210....3L,2002SoPh..210...61H}. Our technique converts the RHESSI time series of a given X-ray source into discrete two-dimensional Fourier components $V(u,v)$ that we call visibilities.
The directness of the sensitivity of visibility-based techniques to an occulted source
is suggested by the relation between visibilities and the spatial profile of a source.  
Let $I(x)$ be the intensity of a two-dimensional source  projected onto the x-axis.  
Then the relationship between the projected intensity and the set of visibilities $V(u)$, whose 
spatial frequencies $u$ are in the x-direction, is given by:
\begin{equation} \label{eqn:vofi}
V(u) =\int I(x) \exp(-2\pi iux) dx
\end{equation}
and therefore
\begin{equation} \label{eqn:iofu}
I(x) = \int V(u) \exp(2\pi iux) du	\ .	
\end{equation} 
This illustrates how the source profile can be reconstructed from a set of visibility measurements.   
Spatially differentiating Equation~\ref{eqn:iofu} yields
\begin{eqnarray}
\frac{dI(x)}{dx}& =& 2 \pi \int iuV(u)\exp(2\pi iux)du \\  
&=&  2 \pi \int V_*(u)\exp(2\pi iux) du \ ,
\end{eqnarray}
which implies that the spatial derivative of the source profile, $I^\prime (x) = dI(x)/dx$, can be reconstructed from a set of modified visibilities, $V_* (u) = iuV(u)$.  
The modified visibilities $V_* (u)$ are thus derived from the measured visibilities, $V(u)$,  by applying a 90 degree phase shift (the factor $i$) and weighting by the spatial frequency $u$. 
$V_* (u)$ is  given by
\begin{equation} \label{eqn:vdash}
V_* (u) =\int I^\prime (x) \exp(-2\pi iux) dx \ .			
\end{equation}
Numerically evaluating Equation~\ref{eqn:vofi} for an unocculted Gaussian source profile results 
in the behavior shown by the black solid curve in  Figure \ref{fig:vis_vs_k}.
 \begin{figure*}
\includegraphics[width=16cm]{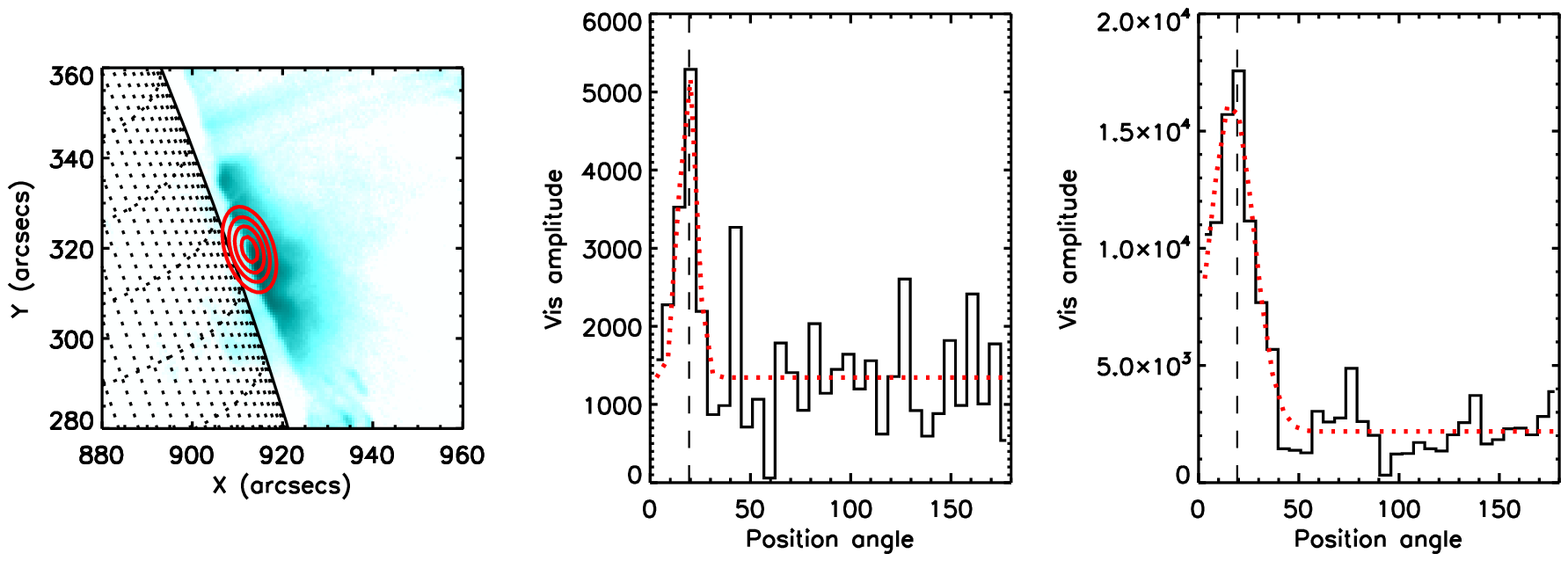}
\caption{Left: AIA 131\AA\ image of SOL2011-10-20T03:25, obtained in the middle of the RHESSI integration. 
The red contours give the 30\%, 50\%, 70\%, and 90\% contours of the RHESSI visibility forward fit at 3-25 keV over a four-minute time interval.
Middle: Visibility amplitude versus position angle for collimator~1; note that we measure position angles CCW from the W~limb in this case.
The dashed line indicates the position angle of the limb at the flare location. 
Right: same as middle, but for collimator~2. The red dotted line gives a Gaussian fit to the data. 
}
\label{fig:sol202-04-04vis}
\end{figure*}
As a smaller source becomes over-resolved at high spatial frequencies, the visibility amplitudes decrease rapidly.  
For a smooth source that is occulted abruptly at $x=x_0$, $I^\prime (x)$ has a sharp peak -- approaching a delta function for an arbitrarily sharp limb -- and evaluating Equation~\ref{eqn:vdash} yields $V_* (u) =  \exp(-2\pi iux_0)$, from which $V(u) = (1/iu) \times \exp(2\pi iux_0)$.  
In other words, at high spatial frequencies we find a contribution to the visibility amplitude that decreases as 1/(spatial frequency).  
This is confirmed numerically in Figure~\ref{fig:vis_vs_k} for various fractional occultations of a Gaussian source (colored solid lines).  
Note that the phase of 
this visibility depends on the location of the occulted edge of the source.
As Figure~\ref{fig:vis_vs_k} suggests, the presence of an occulted edge to an otherwise smooth source results in visibilities with significantly enhanced amplitudes at the higher spatial frequencies.  
However this enhancement applies only to visibilities whose spatial frequencies are orthogonal to the occulted edge.  
Thus the direct signature of occultation is a spike in the amplitudes of visibilities at high spatial 
frequencies, when the direction of these visibilities approaches orthogonality to the occulted edge of 
the source. This is illustrated in the right-hand part of Figure \ref{fig:vis_vs_k}, where the visibility amplitudes as a function of position angle for two sources with 12 arcsec FWHM and 6 arcsec FWHM that are occulted at an angle of 20 degrees are shown. Note how the width of the peak in amplitude depends on the source size. 
To summarize, visibility analysis allows us to distinguish the sharp, precisely oriented image edge expected from limb occultation, from other spatial structures intrinsic to the source.
\begin{figure*}[h!]
\includegraphics[width=14cm]{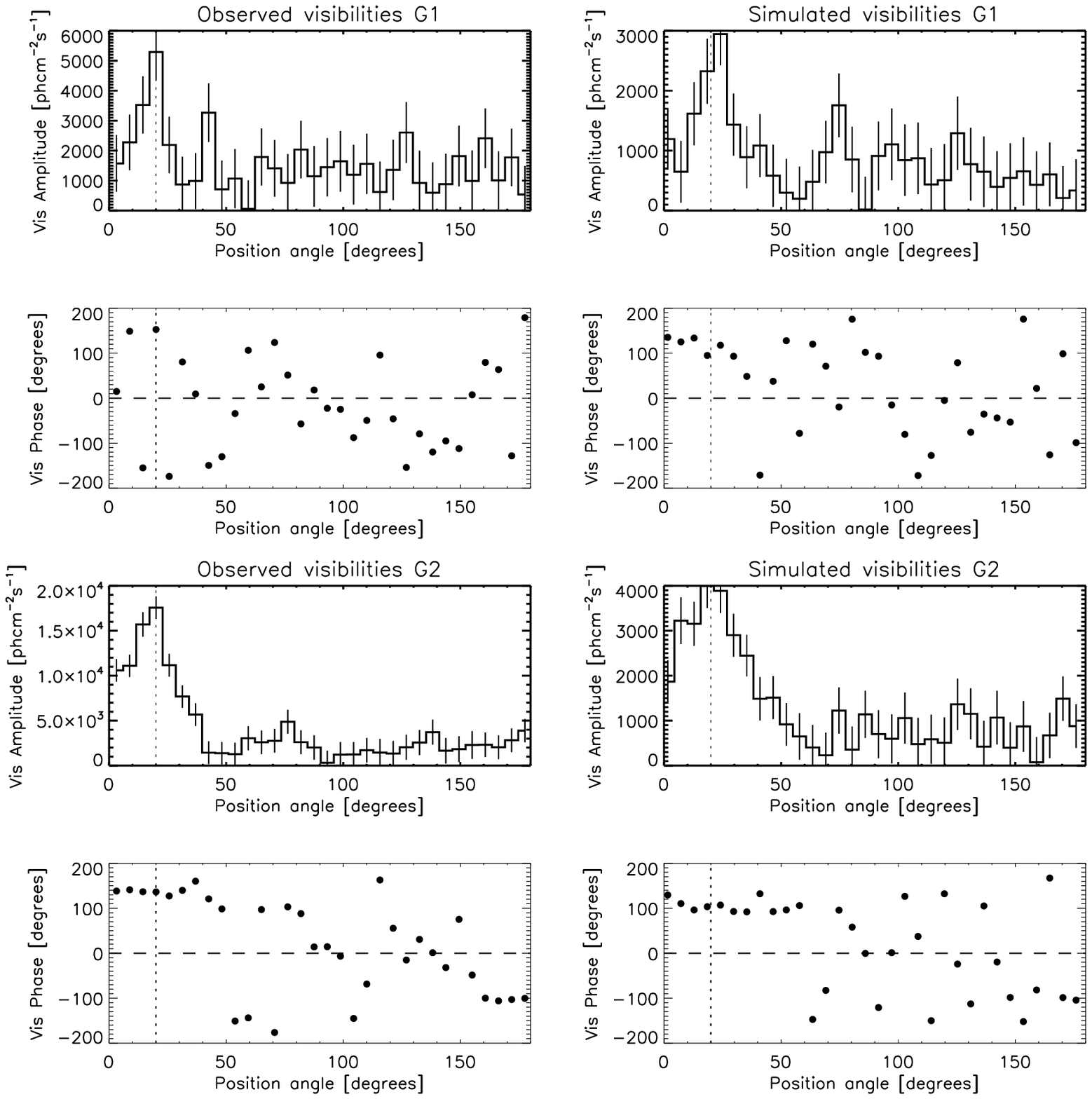}
\caption{Comparison of observed visibility amplitudes and phases from grids 1 and 2 with simulations of a source with 12 arcsec FWHM that is 50\% occulted. Left: observed data. Right: simulations. Top to bottom: Amplitudes and phases of collimator~1 (G1), amplitudes and phases of collimator~2 (G2).}
\label{fig:simulations}
\end{figure*}
For this we have several tests: 
\begin{enumerate}
\item There must be a peak in the high spatial frequency visibility amplitude as a function of position angle
\item The peak must occur at the position angle corresponding to the limb orientation
\item The peak should have an amplitude consistent with the expectation from the total visible source flux and the occultation fraction, while the amplitude of visibilities at other angles must show no such behaviour (compare Figure \ref{fig:vis_vs_k}). 
\item The peak should have width consistent with the source size and collimator pitch (see Figure \ref{fig:vis_vs_k})
\item  The ratios of the peaks in different sub-collimators should be consistent with their relative spatial frequencies
\end{enumerate}
These tests embody model-dependence and are not conclusive; for example if the actual solar limb had a tilt just on the size scale of the source, this would confuse Test 2.
\section{Observations}\label{sec:obs}
\begin{deluxetable*}{lllll}
\tablecaption{Comparison of observed visibility amplitude properties of collimators 1 and 2  with expectation for a 12 arcsec FWHM source at 50\% occultation.\label{tab:obs}}
\tablehead{\colhead{Item} & \colhead{Test} & \colhead{Expectation} & \colhead{Observed} & \colhead{Pass/fail}} 
\startdata
1 & Existence of peaks & Both G1 and G2 & Yes &  {\color{green}pass} \\
2 & Limb angle vs position of peak & identical & 19.3$^\circ$ vs 20$^\circ$ & {\color{green}pass} \\
3 & Vis amplitude in G1 & $\sim 8800$ ph $\mathrm{cm^{-2}s^{-1}}$ & $5377\pm 887$ ph $\mathrm{cm^{-2}s^{-1}}$& possible\\
3 & Vis amplitude in G2 & $\sim 1.7\times 10^4$ ph $\mathrm{cm^{-2}s^{-1}}$& $(1.6\pm 0.1)\times 10^{4}$ ph $\mathrm{cm^{-2}s^{-1}}$& {\color{green}{pass}} \\
4& Width of peak in G1& 20 deg & $9.78\pm 2.20$ deg & possible \\
4 & Width of peak in  G2 & 34 deg & $25.5\pm 2.02$ deg &  possible \\
5 & Ratio of max amplitude G1/G2 & 0.58 & $0.33\pm 0.25$ & {\color{green}{pass}} \\
\enddata
\end{deluxetable*}

Candidate flares for this application must not only be appropriately located and occulted, but their unocculted  emission must be large enough to provide sufficient counting statistics for RHESSI.  
Note the trade-off with the depth of occultation, which results in a larger visibility enhancement (see Figure~\ref{fig:vis_vs_k}) but also lower observable emission.
We have found an ideal candidate, SOL2011-10-20T03:25 (M1.7). Images from SDO/AIA at 131 \AA\ suggest that the flare source was indeed occulted. In addition, RHESSI's detector 2 was performing well enough for its data to be used for this kind of analysis, hence we can use the modulation of G2 as an independent measure.
As a feasibility demonstration we have analyzed the visibilities over an energy band of 3-25 keV, and over the four-minute time interval from 03:15:00 -- 03:19:00~UT, chosen to maximize the signal-to-noise ratio.  
RHESSI's G1, with a spatial resolution of 2.26$''$ (FWHM), and G2, with a resolution of 3.92$''$, have the highest spatial frequencies and so are best suited to observe the spike in visibility amplitude described in Section~\ref{sec:theory}. Figure~\ref{fig:sol202-04-04vis} shows a AIA 131 \AA\ image overlaid with the RHESSI source. Because the signature of the occulted edge of the source is limited to a small azimuthal range for the highest resolution subcollimators, images made with ``traditional'' imaging algorithms cannot be readily interpreted in terms of the edge location and properties. In particular, none of the traditional imaging algorithms use an image basis appropriately optimized for an occulted source, but use some form of circular or Gaussian symmetry for the reconstruction. This is why we must use visibilities directly. 
The visibility amplitudes as a function of position angle for collimators~1 and~2 (G1 and G2) are also shown in Figure \ref{fig:sol202-04-04vis}. 

We have applied the tests described in Section \ref{sec:theory} to these observations.
The first and most important test is that there be distinct peaks in both G1 and G2, at the correct location and well in excess of the uncertainty estimates derived from the point-by-point measurements of the complex amplitudes. The data clearly pass this test. Unfortunately the other tests have some ambiguities and uncertainties. Note that the expectations for these tests must refer to a model that may be inappropriate in some details. In particular, we do not know the actual width of the source nor the fraction by which it is occulted.
A full image made via the visibility forward-fitting technique gives a FWHM of 11$''$. Hence, for comparison, an occultation of 50\% and a circular Gaussian source of FWHM 12 arcsec (giving a lower limit of the amplitude peak width) was used. Compared with this model, the observed peak is narrower (found by fitting a Gaussian to the observed amplitudes, red line in Figure \ref{fig:sol202-04-04vis}) than expected. The ratio of the amplitudes is also lower than it should be, though has quite large uncertainties within which it agrees with the expectation. Note that the latter only depends on the relative spatial frequencies of the grids used and not on the source properties. The results of the tests are summarized in Table~\ref{tab:obs}. In addition to these tests we used the RHESSI simulation software (part of the Solar Software) to calculate the expected visibility amplitudes and phases in a more realistic way. The map center for the reconstructed visiblities was set to [908.8, 318.6] arcsec, the estimated location of the limb, the same as was used for the actual analysis. 
The resulting visibility amplitudes and phases for G1 and G2 are compared with the observed data in Figure \ref{fig:simulations}. The simulations show a peak at the same location and with similar width as the observed data and also the phases agree well. In addition, the visibility amplitudes at position angles not associated with the limb display fluctuations of the same order as the observed visibility amplitudes, suggesting that these are predominantly due to statistical uncertainties, rather than source substructure. Finally, we have to exclude the possibility that the source is not occulted but thin and elongated along the limb, which would result in a similar visibility amplitude behaviour. As can be see in Figure \ref{fig:sol202-04-04vis}, the visibility forward fit suggests an elongated source in the direction of the limb that would be consistent with a non-occulted elliptical source with 15 arcsec FWHM large axis and a 8 arcsec FWHM short axis. In Figure \ref{fig:vis_vs_k} the observed relative amplitudes from the visibilities along the direction of the limb and perpendicular to the limb are shown. The data are clearly inconsistent with a non-occulted source, as the amplitudes in G1 and G2 are several orders of magnitude above the expectation for a non-occulted source and for an 8 arcsec source extent perpendicular to the limb, the difference would be even more pronounced.

In summary the observations clearly pass tests number 1, 2, and 5, which are independent of the source morphology and occultation fraction. The results of tests 3 and 4 are a bit ambiguous, as these depend on a model of the true source whose morphology is not known. In addition, simulations show a similar behaviour as the observations, and we can exclude a thin, non-occulted source parallel to the limb. Thus we interpret the observed spike in amplitude as reflecting the limb edge and analyze its complex amplitudes to derive the limb height in the following.
\begin{figure*}
\includegraphics[width=14cm]{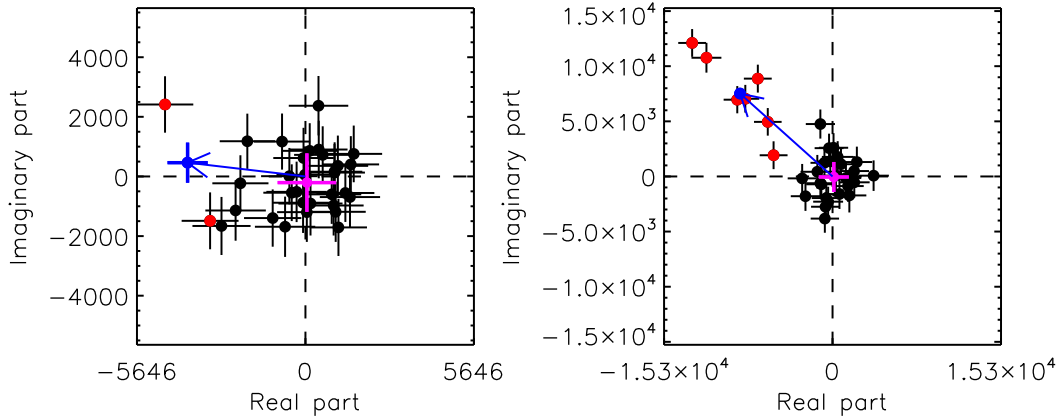}
\caption{Observed visibilities in G1 (left) and G2 (right) in the complex plane. The red dots give the presumed limb signal with an amplitude larger than 3$\sigma$ of its uncertainty. The purple point gives the vectorially added background visibilities. The blue point and arrow give the phase of the limb.}
\label{fig:complex}
\end{figure*}

\section{Observed and predicted limb height} \label{sec:height}
\subsection{Limb height from visibility analysis}\label{sec:visobs}
For a given subcollimator, the information on the radial position $R_{L}$ of the limb is contained in the phase $\phi$ of the complex visibility, where 
\begin{equation} \label{eq:height}
R_{L}=C_M+\frac{(\phi_L -\pi/2)}{2\pi}\times p,
\end{equation}
where $C_M=\sqrt{x_c^2+y_c^2}$ is the position of the map-center used for the visibility calculation, and $p$ is the pitch of the collimator that is used for the analysis (4.52 arcsec for G1 and 7.84 arcsec for G2 which are used here). 
The location of the map center is determined by the metrology of the RHESSI grid optics (essentially, the RHESSI plate scale). Figure \ref{fig:complex} shows the visbilities of G1 and G2 in the complex plane, the basis of the analysis. The visibility phase is the angle measured counter-clockwise from the positive x-axis, the amplitude is the radial distance from the origin of the coordinate system. The visibilities that are dominated by the limb signal were identified as the points around the peak with an amplitude larger than its 3$\sigma$ uncertainty. They are summed up vectorially to give the phase that represents the limb signal. From this, the vectorial sum of all other visibilities (``background") was subtracted. The resulting phase was then used in Equation~\ref{eq:height} to calculate the radial limb position. 
The mean limb location, derived in this way from the observed phase information, is given in Table~\ref{tab:results}.
We show two sets of errors: the formal errors of the visibility fit, based explicitly upon the counting statistics, and a second set estimating the systematic error. The former is derived from the uncertainty of the phase measurement. Note that, even though the resolution of the grids is limited to a few arcseconds, the visibility phases can be measured with much higher accuracy (compare Figure \ref{fig:complex}). The systematic uncertainty consists of three components. The first one uses the empirical scatter of the significant phase measurements at our binning, i.e. the standard deviation of the visibility phases associated with the limb (red points in Figure \ref{fig:complex}).
For the weighted mean we also incorporate two possible instrumental uncertainties: the length of the metering tube that holds the grids, and the accuracy of the Sun-center determination from the solar aspect system. For the former we adopt a 3$\sigma$ error of 0.026$''$ \citep{2003SPIE.4853...41Z}, and for the latter, as an upper limit, 0.15$"$ \citep{2002SoPh..210...87F}. 

\begin{deluxetable*}{ lll }
\tablecaption{Mean limb location for SOL2011-10-20T:03:25, at 1 AU, integrated over photon energy range of 3-25 keV. The uncertainties of the measurement are given as statistical uncertainties from the visibility fit plus systematic uncertainties. \label{tab:results}}
\tablehead{\colhead{Collimator} & \colhead{FWHM ($''$)} & \colhead{Result ($''$)}}            
\startdata
G1 & 2.26 & 960.10 $\pm 0.41 \pm 0.46$ \\
G2 & 3.92 & 960.11 $\pm 0.16 \pm0.21$ \\
Weighted mean (G1+G2) & & $960.11 \pm 0.15 \pm 0.29$ \\
Prediction & & 959.94 (10 keV) \\
\enddata
\end{deluxetable*}

\subsection{Predictions}
\begin{figure}[!]
\includegraphics[width=0.5\textwidth]{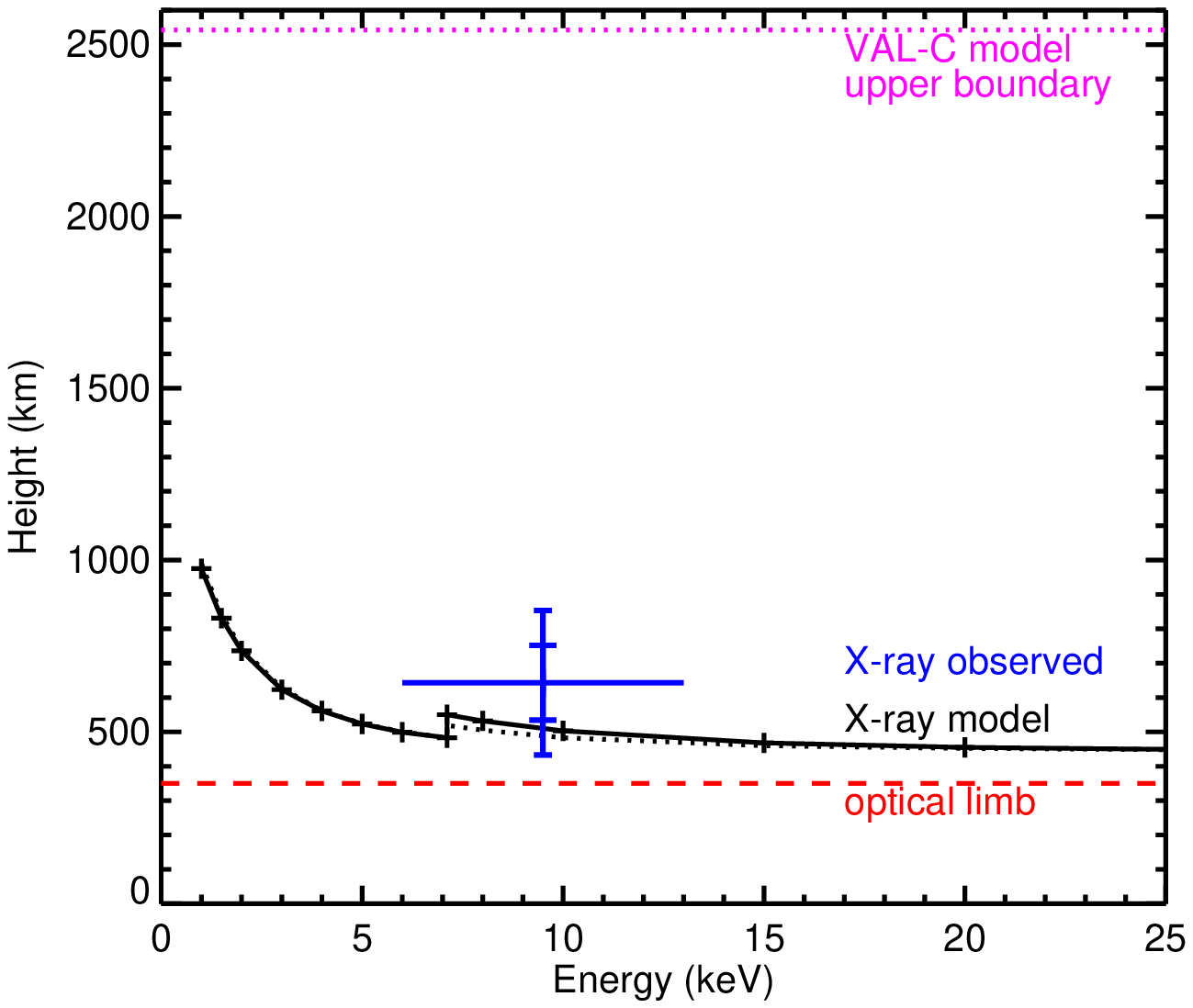}
\caption{Prediction of the occultation height in the VAL-C model atmosphere  \citep{1981ApJS...45..635V} as a function of X-ray energy, making use of abundances from \cite{2011SoPh..268..255C}.
The dotted line shows the same calculation for coronal abundances \citep{2012ApJ...755...33S}. 
The dashed line gives a standard estimate for the height of the optical limb. Our measurement and its uncertainties (statistical and systematic) are given as the blue cross, which is centered at the peak of the RHESSI count spectrum. The horizontal bar indicates the FWHM of the count spectrum peak, which is significantly narrower than the energy range used to make visibilites (3-25 keV).
These estimates assume neutral plasma and spherical symmetry. 
}
\label{fig:valc_height}
\end{figure}

Figure \ref{fig:valc_height} displays our result in terms of height of the observed limb above the photosphere in comparison with the height of the visible limb.
Given the fundamental difference between an X-ray absorption signature of the limb location, its visible location, and the known wavelength-dependence of the solar radius \citep[e.g.][]{2015ApJ...812...91R}, our results are within expectation regarding the difference between X-ray limb height and optical limb.
We also compare the observed limb height with the predicted X-ray limb height from a model atmosphere. For this first result we compare with 
the standard VAL-C semi-empirical model atmosphere \citep{1981ApJS...45..635V}.
We have calculated the X-ray opacity of the photospheric plasma from weighted means of the mass attenuation coefficients of the contributing elements (H, He, C, N, O, and Fe), incorporating both the photoelectric effect and electron (Compton) scattering.
Note that this approach omits any fine structure associated with flows (leading to corrugations), and sphericity.
The predicted limb height varies with photon energy, as shown in Figure~\ref{fig:valc_height} for the photospheric solar elemental abundances of \cite{2011SoPh..268..255C} and lies below the observed height with a 1$\sigma$ significance (see Section \ref{sec:discussion}).
The electron scattering just depends upon the total line-of-sight electron density, including atomic electrons, at X-ray photon energies.
The photoeffect, on the other hand, depends upon the bound-state populations of different ionic states, and requires a detailed model.
We have assumed negligible ionization in the photosphere, and spherical symmetry, to determine the impact factor for 1/e attenuation of the occulted source.
Note that the predicted height only varies slowly with photon energy above the Fe K-edge (visible as discontinuity at 7 keV in Figure~\ref{fig:valc_height}), because of the dominance of electron scattering for X-ray opacity in the photosphere.
Nevertheless the Fe feature does reveal itself significantly, so that with better observations one could obtain a clean determination of a mean Fe abundance in this way.
\section{Discussion}\label{sec:discussion}
The technique we have explored in analyzing the RHESSI observations of SOL2011-10-20T03:25 (M1.7) 
has a particularly simple physical interpretation in terms of the actual mass distribution with height in the atmosphere.
Despite this, and despite the excellent opportunity that the RHESSI rotating modulation collimators afford, we regard this first result as a tentative one.
Partly this is because of the difficulty of finding suitable occulted flares bright enough for this analysis, but mainly it is for systematic reasons that we discuss here.
 
Several factors contribute, the most important in this case being the relative locations of the pointing axis (a few arcminutes west of Sun center), Sun center, and the flare location. The measurement of visibilities whose orientation matches the direction of a line from the pointing axis to the limb source can be compromised by incomplete phase sampling \citep[Figure 4 of][]{2002SoPh..210...61H}. This factor could be alleviated by non-uniform pointing variations (at the relevant roll angle) on a scale of a few arcseconds. Unfortunately in this case, RHESSI's pointing was too stable.
For visibilities tangent to the limb, there is no problem because of the rapid variation of the modulation.
Other lesser problems with the data include the simple problem of counting statistics: only a small fraction of the data apply to just the visibility components needed.
Finally, the theoretical visibility pattern depends to a certain extent on unknown fine-scale image structure that may compete with the sharp edge expected for the occultation. 

Why is the measured X-ray limb location above its predicted height?
Of course, the prediction requires model assumptions. 
In particular our prediction ignores the presence of physical corrugations of the limb \citep{1969SoPh....9..317S,1997ApJ...484..960G}.
The structures would need to be in excess of the hydrostatic scale height of about 100~km in the photosphere, and on small angular scales.
The likeliest sources for such corrugations would be p-mode or convective flows in the photosphere \citep{2000Natur.405..544K} or some lower-atmosphere counterpart of the spicule structure seen in the chromosphere \citep[e.g.][]{1993ApJ...403..426E,2002ApJ...575.1104A}.
We think it unlikely that they have an influence on the model or observations, based on present knowledge, but look forward to higher-resolution optical observations and to more definitive modeling.

Other issues have to do with possible time variability of the limb location, as well as the physical nature of the particular patch of the Sun involved in our single determination; a quiet-Sun model such as VAL-C may simply be inappropriate.
Such a local interpretation might also help to resolve the unexpectedly low altitude found for the hard X-ray sources in a single well-observed flare on the visible hemisphere \citep{2012ApJ...753L..26M}.
The Sun itself may be aspherical, but we note that the observed oblateness term is not large enough to be relevant here \citep{2008Sci...322..560F}.

\section{Conclusion}
The results presented here, to our knowledge the first X-ray measurements of the solar radius, offer great promise.
In spite of the caveats discussed above, we are optimistic that this technique can generate a rigorous and independent measurement of the true solar radius: the X-ray measurements provide a means to determine the actual mass distribution rather than the more model-dependent distribution of opacity
that underlies classical optical techniques.
The result (Table~\ref{tab:results}) does not quite match expectations from the standard VAL-C model. The discrepancy between observation and prediction, at about 1$\sigma$ significance, could imply errors in our use of the VAL-C model, for example by application to an area of the Sun near a major active region.
Corrugations of the atmosphere at these temperature-minimum heights could also play a role.
The technique is sound enough for us to encourage further exploration of the RHESSI database for this purpose, with the ultimate goal of establishing better solar radius determinations this way.
If this becomes possible, improved observations using the X-ray technique could in principle help determine solar iron abundance, taking advantage of the effect of the K-edge visible in Figure \ref{fig:valc_height}. \\ \\

\acknowledgments
This work was supported by Swiss National Science Foundation (200021-140308), by
the European Commission through HESPE (FP7-SPACE-2010-263086), and through NASA  Contract NAS5-98033 for RHESSI. We thank the anonymous referee for helpful comments and suggestions.
\bibliography{xray_limb}
\end{document}